\documentclass[a4paper,nofootinbib,twocolumn]{revtex4-1} 
\pdfoutput=1
\usepackage{graphicx}  
\usepackage{bbold}  
\usepackage{mathtools}
\usepackage{amsmath}
\usepackage{amsfonts} 
\usepackage{amssymb}
\usepackage{amsbsy}
\usepackage{amsfonts}
\usepackage{physics} 
\usepackage{bbold}
\usepackage{slashed}  
\usepackage{color}
\usepackage{tablestyles} 
\usepackage{tabularx}
\usepackage{hyperref} 
\usepackage{orcidlink}
\usepackage{bbding} 
\usepackage{ulem}

\definecolor{oucrimsonred}{rgb}{0.6, 0.0, 0.0} 
\definecolor{DarkGray}{gray}{0.4}
\definecolor{forestgreen}{rgb}{0.13,0.35,0.13}
\definecolor{ocre}{HTML}{F16723}

\usepackage{bm}
\usepackage{mathrsfs}

\def\eq#1{{Eq.~(\ref{#1})}}
\def\eqs#1#2{{Eqs.~(\ref{#1})--(\ref{#2})}}

\def\Tr{\mbox{Tr}\,}

\newcommand{\di}{\mbox{d}}

\def\di{\mbox{d}}

\colorlet{grayline}{gray!70}
\definecolor{blueline}{rgb}{0,0.27,0.55}
\definecolor{DarkGray}{gray}{0.4}
\definecolor{Gray}{gray}{0.6}
\definecolor{oucrimsonred}{rgb}{0.6, 0.0, 0.0}
\definecolor{persianblue}{rgb}{0.11, 0.22, 0.73}
\definecolor{forestgreen}{rgb}{0.13,0.35,0.13}
 \hypersetup{colorlinks, citecolor=forestgreen, linkcolor=forestgreen, urlcolor=forestgreen}
\newcommand{\be}{\begin{equation}}
\newcommand{\ee}{\end{equation}}
\newcommand{\bea}{\begin{eqnarray}}
\newcommand{\eea}{\end{eqnarray}}
\newcommand{\nn}{\nonumber}

\newcommand{\CC}{\operatorname{C}}
\newcommand{\BB}{\operatorname{B}}
\newcommand{\hk}{{\bf k}}
\newcommand{\hr}{{\bf r}}
\newcommand{\hn}{{\bf n}}
\newcommand{\hp}{{\bf p}}

\newcommand*\xbar[1]{%
  \hbox{\;%
    \vbox{%
      \hrule height 0.5pt 
      \kern0.5ex
      \hbox{%
        \kern-0.25em
        \ensuremath{#1}%
        \kern-0.07em
      }%
    }%
  }%
} 
\newcommand{\com}[1]{}
\newcommand{\gsim}{\lower.7ex\hbox{$\;\stackrel{\textstyle>}{\sim}\;$}}
\newcommand{\lsim}{\lower.7ex\hbox{$\;\stackrel{\textstyle<}{\sim}\;$}} 

\newcommand{\bc}{\begin{center}}
\newcommand{\ec}{\end{center}}

\begin{document}

\hypersetup{citecolor = forestgreen,
linktoc = section, 
linkcolor = forestgreen, 
urlcolor = forestgreen
}

\title[]{ \Large \color{oucrimsonred} \textbf{ 
  Local vs.\ nonlocal  entanglement in top-quark pairs at the LHC} }

\author{\bf M. Fabbrichesi$^{a\, \orcidlink{0000-0003-1937-3854}}$}
\author{\bf   R. Floreanini$^{a}\, \orcidlink{0000-0002-0424-2707}$}
\author{\bf L. Marzola$^{{c,d\, \orcidlink{0000-0003-2045-1100}}}$}
\affiliation{$^{a}$INFN, Sezione di Trieste, Via Valerio 2, I-34127 Trieste, Italy}
\affiliation{$^{c}$Laboratory of High-Energy and Computational Physics, NICPB, R\"avala 10,  10143 Tallinn, Estonia}
 \affiliation{$^{d}$Institute of Computer Science, University of Tartu, Narva mnt 18, 51009 Tartu, Estonia}

\begin{abstract}
\noindent	We show that the entanglement observed in top-antitop quark spin states at the LHC is local in the energy region close to the production threshold. In contrast, nonlocal entanglement is observed in the central boosted region defined by a top-quark pair invariant mass $m_{t\bar t} > 800$ GeV and scattering angles $\Theta$ satisfying $|\cos \Theta |<0.2$. This makes top-quark pairs a unique laboratory for studying the interplay between entanglement and Bell locality. The locality of entanglement near the production threshold is further supported by a recent CMS analysis, which reports a significance of more than $5\sigma$. We also demonstrate that there exists a kinematic region where the spin states of the top-antitop quark are separable, yet they exhibit non-zero discord and magic.
\end{abstract}

\maketitle

\textbf{Introduction---} Entanglement is a fundamental property of quantum systems~\cite{Horodecki:2009zz,Benatti:2010,Nielsen:2012yss,bruss2019quantum}. In the past few years, it has been an object of study also within particle physics (see, for instance the review article~\cite{Barr:2024djo} and the papers cited therein) and it has been recently observed in the spin states of top-quark pairs created at the LHC~\cite{ATLAS:2023fsd,CMS:2024pts}. 

Bell nonlocality~\cite{Bell:1964,scarani2019bell} is an even more profound property of certain composite quantum systems;
it manifests in the impossibility of reproducing the correlations among the composing subsystems by local, 
hidden variable stochastic models.
It has been observed in particle physics in $B$-meson~\cite{Fabbrichesi:2023idl} and charmonium decays~\cite{Fabbrichesi:2024rec}. 
 
Whereas entanglement  for pure states is equivalent to Bell nonlocality, for more general mixed states this equivalence does not necessarily hold. Indeed, there exist states that, though entangled, remain Bell local. Most of them can be led back to the Werner state~\cite{PhysRevA.40.4277}, which, for the case of the bipartite qubit system---as that formed by pairs of spin-1/2 particles---is a mixture of the singlet and the identity.

Determining whether an entangled state is local or nonlocal is of interest  because  in the former case the quantum system could be well described with a local hidden variable model~\cite{Augusiak_2014}. Moreover, very few explicit examples of local entangled states have been identified in laboratories so far.

In this Letter, we study the properties of the spins state of the top-quark pairs produced at the LHC to show that the entanglement present in the spin correlations can be, depending on the kinematic region, local or nonlocal. We also find that the quantum states which are locally entangled  seem not to be of the Werner type. In addition, we also identify states that are separable and yet exhibit non-vanishing discord and magic. The top-antitop quark system at the LHC thus provides a laboratory where all possible quantum states can be produced and analyzed.

\vskip0.3cm
\textbf{Analytic study---} Entanglement and the violation of the Bell inequality have been extensively studied for the bipartite system formed by top-quark pairs produced at the LHC~\cite{Afik:2020onf,Fabbrichesi:2021npl,Severi:2021cnj,Aguilar-Saavedra:2022uye,Dong:2023xiw,Han:2023fci}. In this Letter we focus on the interplay between entanglement, Bell nonlocality, magic and discord, and examine their variation across the kinematic space of the process. With this goal in mind, we first investigate the related spin correlations with the available analytic tools, useful to gauge what to expect from the collected collider data. 

Pairs of top quarks are produced at the LHC in proton-proton collisions mainly via strong interactions. The spin state of the produced pairs is then reconstructed by means of the angular distributions of the momenta of suitable top-quark decay products, both for the fully leptonic and semi-leptonic channels.

The density matrix describing the spin state of a system formed by two spin-1/2 particles, a bipartite qubit system, can generically be written as
\begin{widetext}
 \begin{equation}
	\label{eq:rho}
		\rho = \frac{1}{4} \Big[
		\mathbb{1}_2\otimes\mathbb{1}_2 
		+ 
		\sum_i \BB_i^+ \, \qty(\sigma_i \otimes \mathbb{1}_2) 
		+ 
		\sum_j \BB_j^- \, \qty(\mathbb{1}_2 \otimes \sigma_j )
		+
		\sum_{i,j} \CC_{ij} \, \qty(\sigma_i \otimes \sigma_j)
		\Big],
	\end{equation}
	\end{widetext}
with $i,j=r, \,n, \,k$, $\sigma_i$ being the Pauli matrices and $\mathbb{1}_2$ the $2\times 2$ identity matrix. The decomposition refers to a right-handed triad, $\{\hn, \hr, \hk\}$ which we choose so that the spin quantization axis is taken along $\hk$, implying $\sigma_k\equiv\sigma_3$. In the top-antitop pair center of mass (CM) frame we have
	\begin{equation}
		\hn = \frac{1}{\sin \Theta }\qty(\hp \times\hk), \quad \hr = \frac{1}{\sin \Theta }\qty(\hp-  \hk \cos \Theta)\,,
	\end{equation}
where $\hk$ is the direction of the momentum of the top quark and $\Theta$ is the related scattering angle. 
We take $\hp\cdot\hk = \cos\Theta$, with $\hp$ being the direction of one of the incoming proton beams. 
The coefficients $\BB_i^+={\rm Tr}[\rho \, (\sigma_i\otimes \mathbb{1}_2)]$ 
and $\BB_i^-={\rm Tr}[\rho \, (\mathbb{1}_2\otimes\sigma_i )]$ give the polarization state of the single particles---averaged over the relevant kinematic distributions, if experimentally reconstructed, or as functions of the kinematic parameters when analytically computed.   
Similarly, the $3\times 3$ real matrix $\CC_{ij}={\rm Tr}[\rho\, (\sigma_i\otimes\sigma_j)]$ encodes the top-antitop quark spin correlations.

\begin{figure}[h!]
\begin{center}
\includegraphics[width=3in]{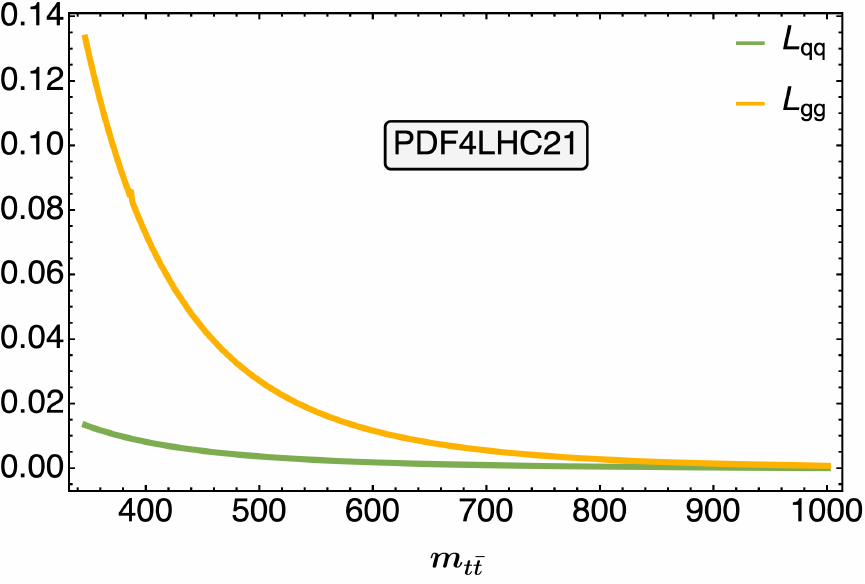}
 \caption{\footnotesize Parton luminosities $L^{gg} (\tau)$ and $L^{q\bar q} (\tau)$ in the range of invariant mass of interest. At threshold the ratio $L^{gg} (\tau)/L^{q\bar q} (\tau)$ is about 9.
\label{fig:pdf} }
\end{center}
\end{figure}

The cross section for the process, summing over the spins of the produced pairs, is given by
\begin{multline}
\frac{\di \sigma}{\di \Omega\, \di m_{t\bar t}} =\\ \frac{\alpha_s^2 \beta_t}{64 \pi^2  m_{t\bar t}^2} \Big\{ L^{gg} (\tau) \, \tilde A^{gg}[m_{t\bar t},\, \Theta]+L^{qq} (\tau)\, \tilde A^{qq}[m_{t\bar t},\, \Theta]  \Big\} \label{x-sec-tt}
\end{multline}
where $\tau = m_{t\bar t}/\sqrt{s}$, $\beta_t=\sqrt{1-4\,m_t^2/m_{t\bar t}^2}$, $\alpha_s=g^2/4 \pi$,  $m_t$ is the top mass, $\sqrt s$ the CM energy and $m_{t\bar t}$ the invariant mass of the top-quark pair. Analytic expressions for the coefficients $\tilde A^{qq}$ and $\tilde A^{gg}$ are given in Ref.~\cite{Fabbrichesi:2022ovb}. The combination of the two channels in \eq{x-sec-tt},  $g+g\to t +\bar t$ and $q + \bar q \to t +\bar t$, is weighted by the respective parton luminosity functions
\bea
L^{gg} (\tau)&=& \frac{2 \tau}{\sqrt{s}} \int_\tau^{1/\tau} \frac{\di z}{z} q_{g} (\tau z) q_{g} \left( \frac{\tau}{z}\right)\quad \text{and}\\
L^{qq} (\tau)&=& \sum_{q=u,d,s}\frac{4 \tau}{\sqrt{s}} \int_\tau^{1/\tau} \frac{\di z}{z} q_{q} (\tau z) q_{\bar q} \left( \frac{\tau}{z}\right)\, ,
\eea
where we indicated with $q_{q,g}(x)$ the parton distribution functions (PDFs). We have used the subset 40 of the ({\sc PDF4LHC21})~\cite{PDF4LHCWorkingGroup:2022cjn} set, with $\sqrt{s}=13$ TeV and  factorization scale $q_0=m_{t\bar t}$, for their numerical evaluation.  Fig.~\ref{fig:pdf} show the values of the parton luminosities in the range of invariant masses of interest.

Because of the symmetries respected by the strong interaction, the polarizations of the top and antitop quark vanish in first approximation. The density matrix describing the spin state of the fermion pairs is then fully determined by the correlation coefficients 
\be
\CC_{ij} [m_{t\bar t},\, \Theta]= \frac{L^{gg} (\tau)\, \tilde C_{ij}^{gg}[m_{t\bar t},\, \Theta]+L^{qq} (\tau)\, \tilde C_{ij}^{qq}[m_{t\bar t},\, \Theta]} {L^{gg}(\tau) \, \tilde A^{gg}[m_{t\bar t},\, \Theta]+L^{qq} (\tau)\, \tilde A^{qq}[m_{t\bar t},\, \Theta]} \, .\label{pdf}
\ee
The $\tilde C_{ij}$ coefficients were first computed in~\cite{Bernreuther:2010ny}; their analytic expressions can be found in~\cite{Fabbrichesi:2022ovb}. Next-to-leading order (NLO) contributions where computed in~\cite{Czakon:2020qbd} (NNLO QCD) and estimated in~\cite{Frederix:2021zsh} (NLO EW+QCD). These corrections mostly affect the correlation coefficients that are small or vanishing. We estimate through direct computation that their overall impact on entanglement and Bell inequality violation
 is at the 3\% level when estimated with respect to the non-vanishing correlations at the LO. Because some of the symmetries present at the LO (like parity, which enforces vanishing $\BB_{i}^{\pm}$ coefficients) are broken at the NLO, the effect of these corrections might be larger on potential asymmetries.

\begin{figure}[h!]
\begin{center}
\includegraphics[width=0.9 \linewidth]{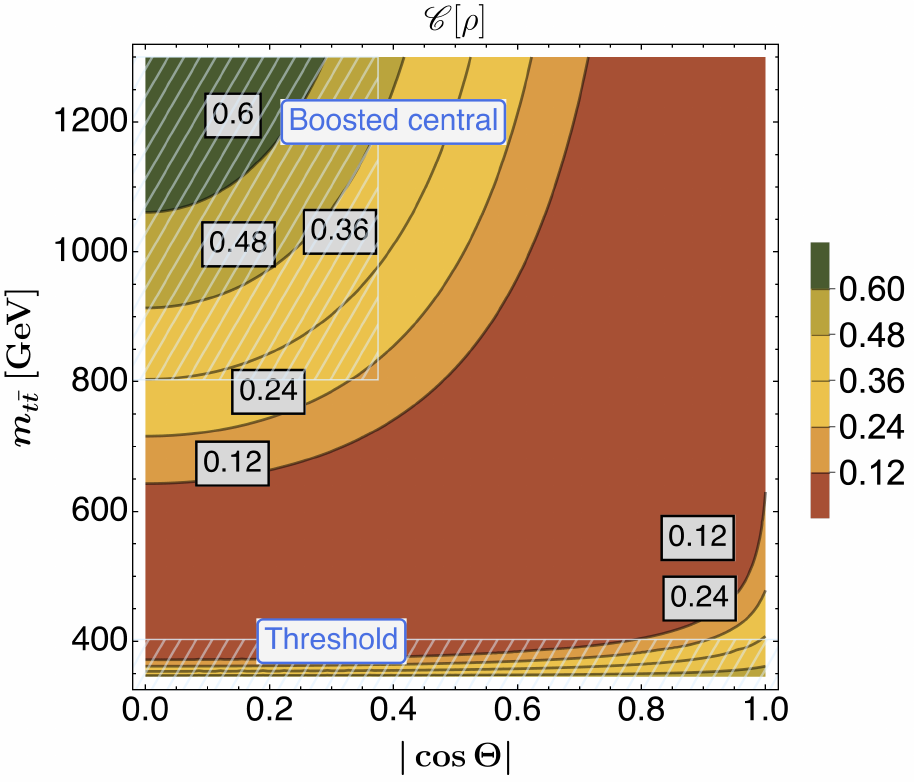}\vskip0.3cm
\includegraphics[width=0.9 \linewidth]{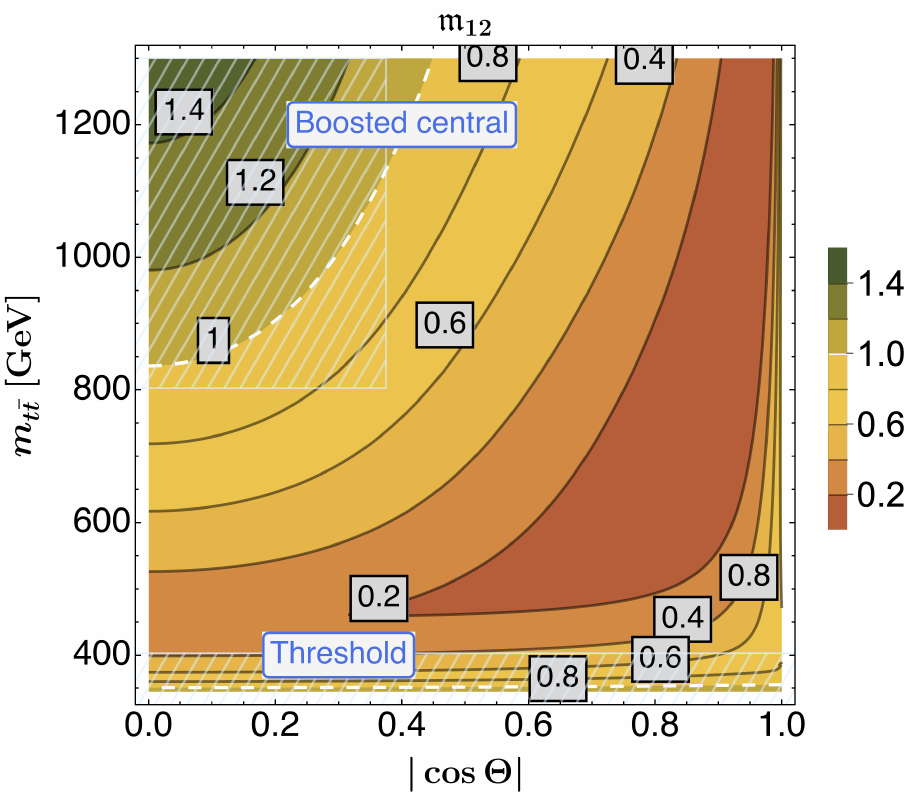}
 \caption{\footnotesize The behavior of the concurrence, $\mathscr{C}[\rho]$, and of the Horodecki parameter $\mathfrak{m}_{12}[\CC]$ over the considered kinematic space. The white, dashed line marks the $\mathfrak{m}_{12}[\CC]=1$ contour, above which the condition for Bell nonlocality is satisfied. The hatched areas denote two of the bins used by the CMS collaboration in their data analysis~\cite{CMS:2024zkc}. 
\label{fig:bin_comb} }
\end{center}
\end{figure}

The entanglement content of top-quark pair spin states can be quantified by means of the concurrence, $\mathscr{C}[\rho]$. This quantity vanishes for separable, unentangled states and reaches its maximal value of 1 for states that are maximally entangled. The concurrence can be analytically computed via the non-negative, auxiliary matrix
$R=\rho \,  (\sigma_y \otimes \sigma_y) \, \rho^* \, (\sigma_y \otimes \sigma_y)$,
where $\rho^*$ is a matrix obtained from the density matrix though the complex conjugation of its entries. The square roots of the eigenvalues of $R$, $r_i$, $i=1,2,3,4$, can be ordered in decreasing order and the concurrence of the state $\rho$ it then given by~\cite{Wootters:PhysRevLett.80.2245}
\begin{equation}
\mathscr{C}[\rho] = \max \big( 0, r_1-r_2-r_3-r_4 \big)\, .
\label{concurrence}
\end{equation}

An alternative entanglement quantifier is the negativity, defined in terms of the eigenvalues $t_{i}$ of the partial transpose  of density matrix in \eq{eq:rho}:
\be
\mathcal{N}[\rho]=\dfrac{1}{2} \left(\sum_{i} |t_{i}|-1\right).
\ee


 \begin{figure}[h!]
 \begin{center}
 \includegraphics[width=0.9 \linewidth]{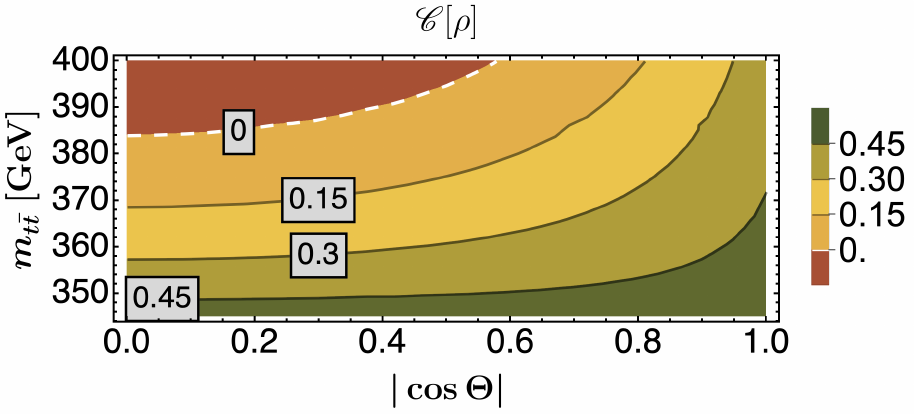}\vskip 0.3cm
\includegraphics[width=0.9 \linewidth]{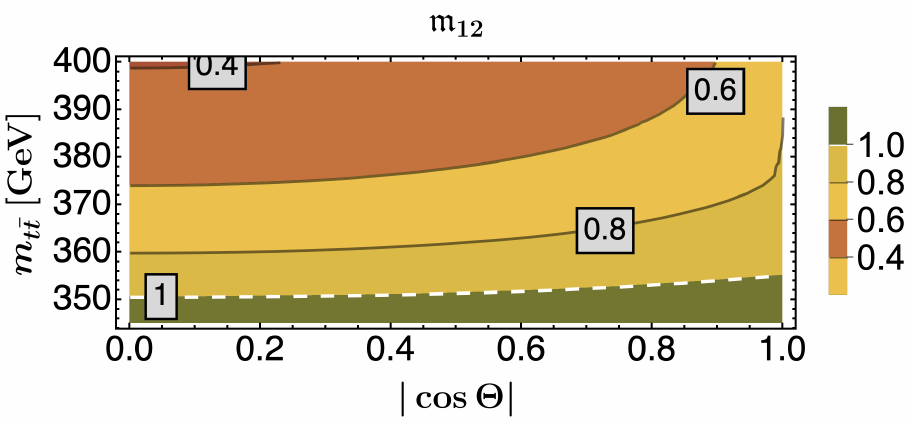}
 \caption{\footnotesize Enlarged view of concurrence $\mathscr{C}[\rho]$ and $\mathfrak{m}_{12}[\CC]$ parameter in the threshold bin defined by the invariant mass $340 < m_{t\bar t} < 400$. The white, dashed lines mark the Horodecki condition for Bell nonlocality, in the lower panel, and a vanishing concurrence in the upper one.
 \label{fig:bin_t} }
 \end{center}
\end{figure}


For the purpose of establishing the presence of Bell nonlocality, it is convenient to introduce the parameter $\mathfrak{m}_{12}[\CC]$, defined as
\be
\mathfrak{m}_{12}[\CC]\equiv m_1 + m_2\, ,
\label{C12}
\ee
in which $m_1$ and $m_2$ are the two largest eigenvalues of the matrix $M=\CC^T \CC$. As shown in Ref.~\cite{HORODECKI1995340}, a two-qubit state $\rho$ with a correlation matrix $C$ satisfying the Horodecki condition
\be
\mathfrak{m}_{12}[\CC] >1\, ,
\label{inequality-test}
\ee
violates the Clauser-Horne-Shimony-Holt (CHSH) formulation of the Bell inequality~\cite{CHSH_1969, scarani2019bell} and, therefore, is Bell nonlocal.  

In addition to the quantities above, the experimental collaborations~\cite{ATLAS:2023fsd,CMS:2024pts} have also  made use of the coefficients
\be
D = \dfrac{1}{3} \Tr \CC \quad \text{and} \quad \tilde{D} =   \dfrac{1}{3} (\CC_{kk}+\CC_{rr}  -\CC_{nn})\,,
\label{D}
\ee
which serve as entanglement witnesses~\cite{Bernreuther:2004jv, Afik:2020onf, Aguilar-Saavedra:2022uye, Baumgart:2012ay}
\be
 D < -\dfrac{1}{3} \quad \text{and} \quad \tilde{D} > \dfrac{1}{3}
\ee
and that can be experimentally reconstructed easily, by measuring the angle between the two final-state charged leptons produced in the decay of the top-quark pair.

Figure~\ref{fig:bin_comb} shows the values of $\mathscr{C}[\rho]$ (top panel) and $\mathfrak{m}_{12}[\CC]$ (bottom panel) over the consider region of the kinematic space. Only in the central region (colored in maroon) $\mathscr{C}[\rho]$ vanishes and the final state of the top-quark pairs is separable---in the remaining regions the spin states of the top-antitop quarks are always entangled. The white, dashed line in the bottom plot shows where the Horodecki condition assumes its threshold value: below the contour the spin states of the top-quark pairs are Bell local. The states in the remaining region are instead Bell nonlocal. The two hatched areas denote two of the bins used in the experimental analysis of Ref.~\cite{CMS:2024zkc}, in which the coefficients of the density matrix are reconstructed in different kinematic limits. Fig.~\ref{fig:bin_t} shows the results of the same analysis in the enlarged kinematic space corresponding to the threshold bin.

Averaging the analytic results over the indicated bins, we find 
\bea
\mathscr{C}[\rho] &=&0.24\pm 0.01\, \,\, (D=-0.49 \pm 0.02)\,,\nn \\
\mathfrak{m}_{12}[\CC] &=& 0.70 \pm 0.02\label{threshold}
\eea
in the threshold bin $340 < m_{t\bar t} < 400$ GeV and $0\leq \cos \Theta\leq 1$.

In order to obtain $\mathfrak{m}_{12}[\CC] >1$ at threshold  we would have to cut the invariant mass at about 350 GeV---which is rather unrealistic if we also want a significant number of events. Further effects such as NLO contributions~\cite{Nason:2025hix} or the presence of top anti-top bound states~\cite{Hagiwara:2008df,Fuks:2021xje,Aguilar-Saavedra:2024mnm,CMS:2025kzt} might here enhance the entanglement.

We also find
\bea
\mathscr{C}[\rho] &=&0.48\pm 0.01\,\,\, (\tilde{D}=0.66\pm0.02)\,, \nn \\
\mathfrak{m}_{12}[\CC] &=& 1.1 \pm 0.03 \label{boosted}
\eea
in a restriction of the boosted central bin defined by $m_{t\bar t} > 800$ GeV and $|\cos \Theta|<0.2$. The strong cut on the scattering angle is to ensure the presence of Bell nonlocality. 

The quoted uncertainties are theoretical in nature and are primarily driven by those associated with NLO corrections. Uncertainties stemming from the input value of the top-quark mass and from the PDFs are at the level of a few per mille.

The results in \eqs{threshold}{boosted} show what to expect from  experimental studies of the process: as a result of the averaging over the respective kinematic regions, the top-quark pairs falling in the  threshold bin are  entangled but in a Bell local state; those in the  central boosted bin are both entangled and in a Bell nonlocal state.

\vskip0.3cm
\textbf{Qualitative features---} As shown in Fig.~\ref{fig:bin_comb}, the entanglement content of the of the $t \bar{t}$-state 
as described by the density matrix in \eq{eq:rho} exhibits a rather complex pattern which depends on the scattering angle $\Theta$, as well as
on the transferred momentum $\beta_t$. Yet, the behavior simplifies at $\Theta=\pi/2$, when the top-quark pair is transversely produced. For this configuration, let us choose the three vectors
$\{\hat{r},\, \hat{k}, \,\hat{n}\}$ to point in the $\{\hat{x},\hat{y},\hat{z}\}$ directions.
In this frame, let us denote by $|\downarrow\rangle$ and $|\uparrow\rangle$ the eigenvectors of the Pauli matrix $\sigma_z$
associated with the eigenvalues $-1$ and $+1$, respectively. Similarly, let 
$|\small{\mp}\rangle=(|\uparrow\rangle \mp |\downarrow\rangle)/\sqrt{2}$  be the eigenvectors of $\sigma_x$, and 
$|\circlearrowleft,\circlearrowright\rangle=(|\uparrow\rangle \mp i|\downarrow\rangle)/\sqrt{2}$ those of $\sigma_y$.

Close to the production threshold, $\beta_t\simeq 0$, the $t \bar{t}$ spin density matrix can be roughly approximated by~\cite{Fabbrichesi:2022ovb}
\be
\rho = p\, \rho^{(-)} + (1-p)\, \rho^{(1)}_{\rm mix}\ ,\qquad 0\leq p \leq 1\ ,
\label{rho-treshold}
\ee
where $\rho^{(-)} = |\psi^{(-)}\rangle\langle\psi^{(-)}|$ is the projector selecting the maximally entangled pure state
\be
|\psi^{(-)}\rangle=\frac{1}{\sqrt{2}}\big( |\downarrow\uparrow\rangle - |\uparrow\downarrow\rangle \big)\ ,
\label{psi-minus}
\ee
while 
\be
\rho^{(1)}_{\rm mix}=\frac{1}{2}\Big( |{\small ++}\rangle\langle {\small ++}| + 
|\small{--}\rangle\langle {\small --}| \Big)\ ,
\ee
is a mixed, separable state. The first contribution comes from gluon-gluon production channel, while the mixed component is due to the quark-antiquark initial state. The relative weight $p$ is regulated by the parton luminosities shown in Fig.~\ref{fig:pdf}. Due to the contribution of the separable mixed state, the entanglement content of the density matrix (\ref{rho-treshold}) is non vanishing only when $p > 1/2$. In addition, for $1/2<p<1/\sqrt{2}$ the entanglement of $\rho$ is not enough to cause the violation of the Bell inequality, thus exhibiting local entanglement.\footnote{Interestingly, the state does not resemble a Werner state~\cite{PhysRevA.40.4277}.}

Differently, at large enough transferred momentum $\beta_t^2 > 0.8$, implying $m_{t\bar t} > 800$ GeV, the $t \bar{t}$ spin density matrix can be
expressed as
\be
\rho = \rho^{(+)} + 2(1-\beta_t^2) \Big( \rho^{(1)}_{\rm mix} + p\, \rho^{(2)}_{\rm mix}- (1+p) \rho^{(+)}\Big) \ ,
\label{rho-large}
\ee
where $\rho^{(+)} = |\psi^{(+)}\rangle\langle\psi^{(+)}|$, with
\be
|\psi^{(+)}\rangle=\frac{1}{\sqrt{2}}\big( |\downarrow\uparrow\rangle + |\uparrow\downarrow\rangle \big)\ ,
\label{psi-plus}
\ee
while 
\be
\rho^{(2)}_{\rm mix}=\frac{1}{2}\Big( |\circlearrowleft\, \circlearrowright\rangle\langle \circlearrowleft\, \circlearrowright| + 
|\circlearrowright\, \circlearrowleft\rangle\langle \circlearrowright\, \circlearrowleft| \Big)
\ee
is a different mixed, separable state. Again, $p$ is the relative weight determined by the parton luminosities.  Close to the limit $\beta_t\to1$, the $t \bar{t}$ polarization state $\rho$ is dominated by the maximally
entangled state $|\psi^{(+)}\rangle$, produced both by gluon-gluon and quark-antiquark initial states. As a result, in this regime, the Bell inequality is maximally violated.

In the intermediate kinematical regime, $0\leq\beta_t^2\leq 0.8$, the $t \bar{t}$ spin state becomes more mixed, leading to the almost complete loss of any quantum correlation. 

These features are only qualitative and hold only in restricted areas of the kinematic space. It is therefore necessary to average over the ranges corresponding to specific bins when comparing the theoretical predictions with the experimental results.

\vskip0.3cm
\textbf{Reinterpretation of CMS data---} The full set of coefficients $\BB^\pm_i$ and $\CC_{ij}$ in \eq{eq:rho} that determine the spin state of an ensemble of  top-quark pairs has been recently published by the CMS experimental collaboration~\cite{CMS:2024zkc}. This work is a treasure trove of useful results. It makes  possible to put to test the analytic estimates presented in the previous section and, to do so, we reinterpret the CMS results in terms of the concurrence and Bell nonlocality observables as shown in Table~\ref{tab:CMS}. The reader should bear in mind that the different choice of the reference triad implies a change of sign in the $k$, $r$ and $n$ directions.

Experimental uncertainties were propagated via Monte Carlo simulation to determine those affecting the observables under consideration. We included correlations among the uncertainties of the $\CC_{ij}$ parameters~\cite{corr} but neglected those with the $\BB^\pm_i$ and $c$ coefficients, as well as those among quantities belonging to different bins which we we do not use.

\begin{table}[h!]
	\centering
	\tablestyle[sansboldbw]
	\begin{tabular}{@{}>{\centering\arraybackslash}p{0.3\linewidth}@{}>{\centering\arraybackslash}p{0.3\linewidth}@{}>{\centering\arraybackslash}p{0.3\linewidth}@{}}
	\theadstart
	\thead&  \thead\textbf{Threshold} &  \thead\textbf{Boosted central} \\
	\tbody
	$\mathscr{C}[\rho]$ & $0.133 \pm 0.055$ & $0.52 \pm 0.06$ \\
	$\mathcal{N}[\rho]$ & $0.067 \pm 0.027$& $0.262\pm0.032$ \\
	$D (\tilde{D})$ & $-0.382  \pm 0.030$ & $(0.662 \pm 0.052)$ \\
	$\mathfrak{m}_{12}[\CC]$ & $0.548 \pm 0.084$ & $1.05 \pm 0.13$ \\
	\tend
	\end{tabular}
	\caption{\label{tab:CMS} Concurrence, negativity and Horodecki parameter as computed from the CMS data pertaining to the two bins: \textbf{Threshold} ($300 < m_{tt} < 400$ GeV) and \textbf{Boosted central} ($m_{tt} > 800$ GeV, $|\cos \Theta| < 0.4$).  The value of the coefficients $D$ and  $\tilde{D}$ are those quoted in~\cite{CMS:2024zkc}.
}
\end{table}

In the threshold region, $m_{t \bar t} < 400$ the significance of the presence of entanglement is only $2.4\sigma$ if the concurrence is used as measure. Nevertheless, one can here make use of the coefficient $D$ (see \eq{D}), as reconstructed from the final lepton relative directions. This has been employed in Refs.~\cite{ATLAS:2023fsd,CMS:2024pts}, and the condition $D<-1/3$, which signals entanglement, has been found to be satisfied with a   significance of about $5\sigma$. Bell locality, on the other hand, is verified with a significance of more than $5\sigma$ by means of the Horodecki condition.

The other region considered corresponds to the boosted central bin, $m_{t \bar t} > 800$ and $|\cos \Theta | < 0.4$. The corresponding kinematic configurations show entanglement with a significance above the $5\sigma$ threshold, though Bell nonlocality cannot be yet asserted. As shown by the analytic results in \eq{boosted}, the bin should be further restricted to $|\cos \Theta | < 0.2$ and the cut on the invariant mass raised in order to reach a better significance.

The CMS analysis also considers a central bin defined by invariant masses $600 < m_{t \bar t} < 800$ GeV. In this bin the top-quark pairs spin states are separable and entanglement  vanishes.


 \begin{figure}[h!]
 \begin{center}
 \includegraphics[width=0.9\linewidth]{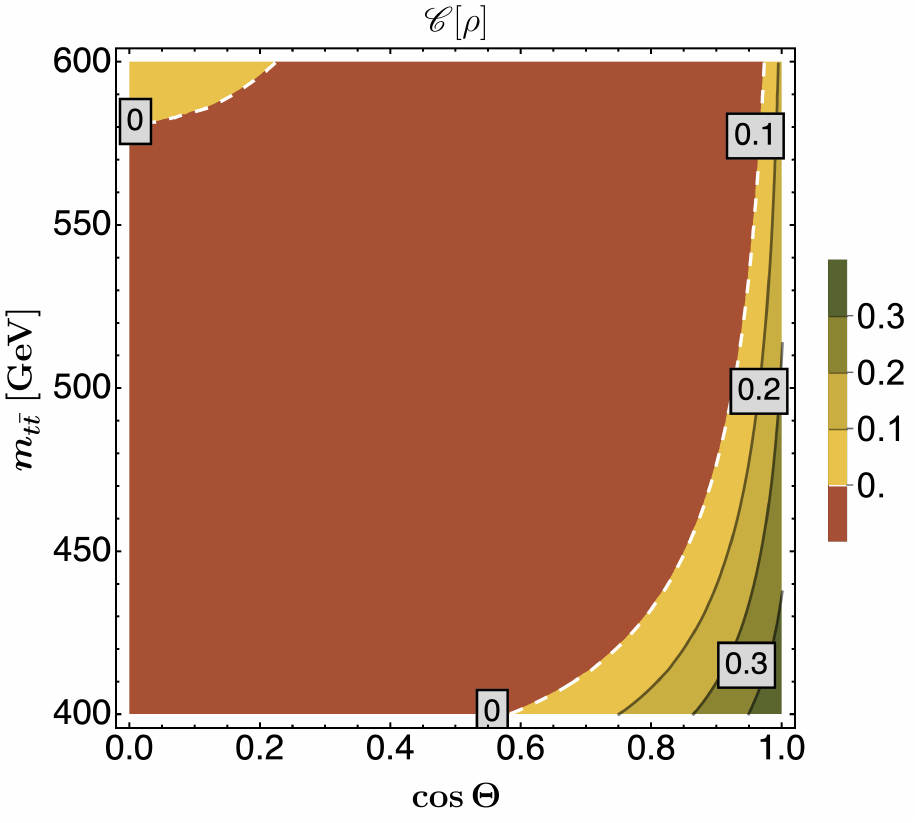} \vskip1cm
 \includegraphics[width=0.9\linewidth]{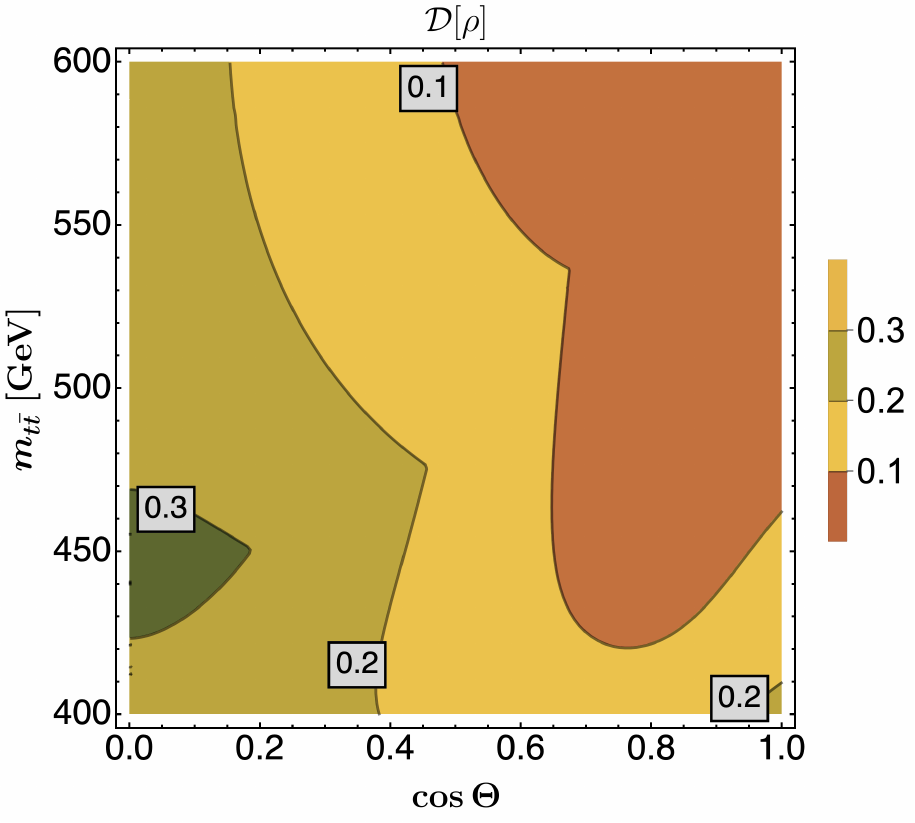}
  \caption{\footnotesize  Concurrence $\mathscr{C}[\rho]$ and discord  $\mathcal{D}[\rho]$ in the intermediate bin $400 < m_{t\bar t}<600 $ GeV. The reddish region between the white dashed-lines,  in the concurrence plot on top, contains separable states for which the concurrence (and the negativity) vanish. On the other hand, the plot below shows non-vanishing values of the discord in the same kinematic region.  
\label{fig:c_d} }
 \end{center}
 \end{figure}

\vskip0.3cm
\textbf{Discord and magic of separable states---} Discord represents the least stringent requirement for a quantum state to be non-classical, as it captures the presence of correlation of quantum origin that not necessarily lead to the presence of entanglement. Technically, it matches the disagreement between two classically equivalent expressions for the mutual information, which generally differ if the involved system is quantum. Given a bipartite qubit system with a density matrix having vanishing coefficients $\BB_i^\pm$, the discord can be computed as
\be
\mathcal{D}[\rho]= 1 - \mathcal{S}[\rho] + h \left( \frac{1+c^{\text{max}}}{2}\right) \geq 0,
\ee
in which $\mathcal{S}[\rho] = - \sum_{i} \lambda_i \log_2 \lambda_i$ is the von Neumann entropy of the top-pairs described by the density matrix $\rho$, with eigenvalues $\lambda_i$. In the equation above, the function 
\be
h(x) = - x \log_2 x - (1-x) \log_2 (1-x)\,,
\ee
depends on $c^{\text{max}}$: the largest of the eigenvalues of the spin correlation matrix $\CC_{ij}$. The discord of the top-quark pair system has been studied in~\cite{Afik:2022dgh,Han:2024ugl} and is noteworthy because it can be used to characterize quantum states that are separable but not classical.

\begin{figure}[h!]
 \begin{center}
  \includegraphics[width=0.9 \linewidth]{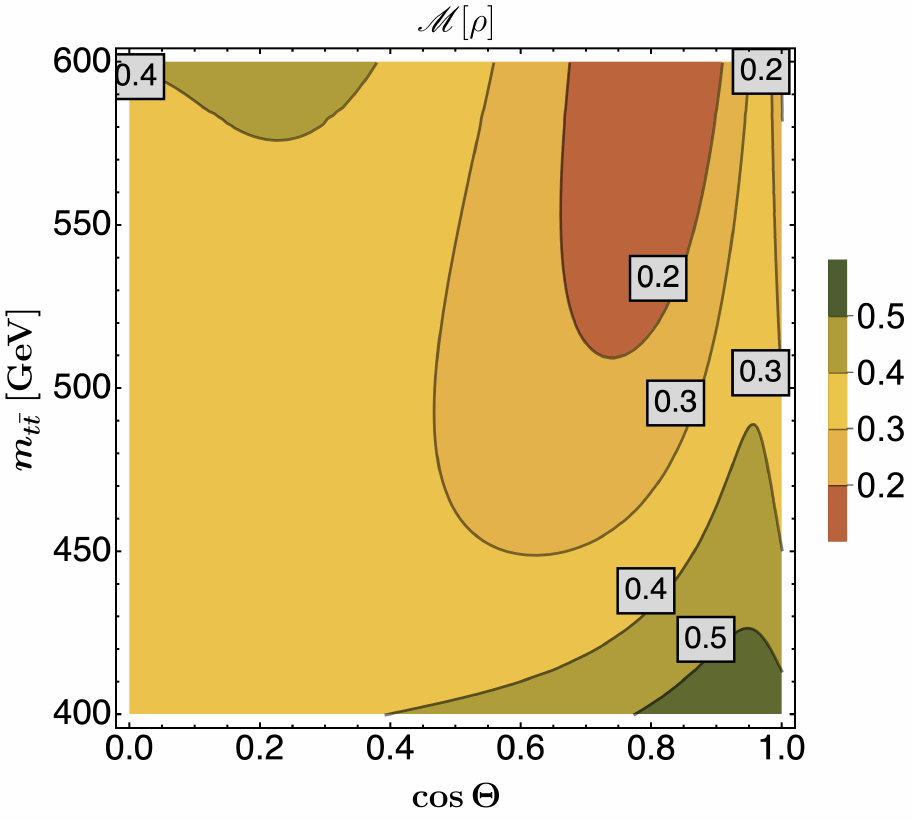}
 \caption{\footnotesize  Magic $\mathscr{M}[\rho]$ of the top-quark pairs produced at the LHC  in the intermediate region $400 < m_{t\bar t}<600 $ GeV.   
\label{fig:c_m} }
 \end{center}
 \end{figure}

Another way to characterize a quantum state is through its magic content. Magic is the resource that allows for quantum computer to outperform their classical counterparts, hence magic states are of importance within quantum information theory because they are better suited for quantum computations  than maximally entangled ones. While the role of this observable in high-energy physics is less clear, we include here its treatment for the sake of completeness. In the case of the top-quark pairs production at the LHC, this observable can be written as a function of the correlation coefficients $\CC_{ij}$ as~\cite{White:2024nuc} 
\be
\mathscr{M}[\rho]= - \log_2 \left[ \frac{1 + \sum_{i,j=1}^{3} \CC_{ij}^4}{1 + \sum_{i,j=1}^{3} \CC_{ij}^2}\right].
\ee
In parts of the intermediate kinematic region defined by the cut $400 < m_{t\bar t} < 600$ GeV, the concurrence (and the negativity as well) vanish, as shown in the first panel of  Fig.~\ref{fig:c_d}. This indicates that, in the corresponding kinematic configurations, the top-antitop quark pairs are produced in states that are, by definition, separable. Nevertheless, we still observe the presence of genuine quantum correlations in the system, as testified by the non-vanish values of the discord. Although separable, states of the top-quark pairs produced in this kinematic regions are 
not classical. The behavior of discord on the considered kinematic space is shown in the second panel of Fig.~\ref{fig:c_d}, as a function of the top-quark pair invariant mass and center-of-mass scattering angle. Similarly, the magic content of the top-quark pair states is shown in Fig.~\ref{fig:c_m} as a function of the same quantities. Both these plots use  analytic results.

\begin{table}[h!]
	\centering
	\tablestyle[sansboldbw]
	\begin{tabular}{@{}>{\centering\arraybackslash}p{0.3\linewidth}@{}>{\centering\arraybackslash}p{0.3\linewidth}@{}}
	\theadstart
	\thead&  \thead\textbf{Intermediate} \\
	\tbody
	$\mathscr{C}[\rho]$ & $0.0 \pm 0.02$ \\
	$\mathcal{N}[\rho]$ & $0.0 \pm 0.02$ \\
	$\mathcal{D}[\rho]$ & $0.097\pm0.013$  \\
	$\mathscr{M}[\rho]$ & $0.292\pm 0.019$ \\
	\tend
	\end{tabular}
	\caption{\label{tab:CMS2}Concurrence, negativity,  discord and magic as computed from the CMS data pertaining to the  \textbf{Intermediate} bin  ($400 < m_{tt} < 600$ GeV). 
}
\end{table}

CMS data~\cite{CMS:2024zkc} can be reinterpreted also in this case to confirm that in the intermediate kinematic region the states are separable but not purely classical, and to find the value of their discord and magic as presented in Table \ref{tab:CMS2}.

\vskip0.3cm
\textbf{Outlook---} The study of quantum entanglement at particle colliders offers the possibility of capturing various aspects of quantum correlations in a single setting  by suitably varying the kinematical configuration. In particular, we have shown that the spin state of top-quark pairs produced at the LHC give rise to a great variety of bipartite qubit states whose properties are determined by the kinematic variables regulating the production process, as well as by the parton luminosities that modulate the contributions of gluon and quark initial states. Close to the top-antitop production threshold,  the spin state of the system contains entanglement, although it does not show any Bell nonlocal correlation. Consequently, the system could be here described by a local hidden variable model.\footnote{As a side remark, let us mention that such local entangled states may, nevertheless,
show some hidden form of nonlocality that can be made explicit through additional
activation processes~\cite{scarani2019bell}, as {\it distillation} through many-copy state preparation~\cite{Peres-PhysRevA.54.2685},
{\it filtering}~\cite{Popescu-PhysRevLett.74.2619,Gisin-1996151} or {\it superactivation}~\cite{Palazuelos-PhysRevLett.109.190401}. None of these protocols are presently implementable at colliders.}
True nonlocal entanglement can  be found in the boosted central region, where it must be further sought to  improve on the current results which are yet unable to establish the presence of Bell nonlocality. In the intermediate kinematic region, the top-quark pair states are separable but exhibit non-vanishing discord (and magic), thus signaling their non-classical nature.

\vskip0.3cm
\textit{Acknowledgements---} {\small MF thanks P.\ Caban and M.\ Pinamonti for discussions.
LM is supported by the Estonian Research Council under the RVTT3, TK202 and PRG1884 grants.}
\vskip1cm
\small
\bibliographystyle{JHEP}   
\bibliography{local.bib} 

\end{document}